\documentclass[conference]{IEEEtran}
\IEEEoverridecommandlockouts
\usepackage[a4paper, total={184mm,239mm}]{geometry}
\usepackage{cite}
\usepackage{amsmath,amssymb,amsfonts}
\usepackage{interval}
\usepackage{caption}
\usepackage{subfig}
\usepackage{algorithmic}
\usepackage{graphicx}
\usepackage{textcomp}
\usepackage{tabularray}
\usepackage{xcolor}
\usepackage[hyphens]{url}
\usepackage{fancyhdr}
\usepackage{hyperref}
\usepackage[capitalise]{cleveref}
\usepackage[acronym]{glossaries}
\usepackage{adjustbox}
\usepackage{siunitx}

\def\BibTeX{{\rm B\kern-.05em{\sc i\kern-.025em b}\kern-.08em
    T\kern-.1667em\lower.7ex\hbox{E}\kern-.125emX}}
\begin{document}

\bstctlcite{IEEEexample:BSTcontrol}

\title{Optimizing Offload Performance in Heterogeneous MPSoCs}
\author{\IEEEauthorblockN{Luca Colagrande}
\IEEEauthorblockA{\textit{Integrated Systems Laboratory, IIS}\\
ETH Zürich, Switzerland \\
colluca@iis.ee.ethz.ch}
\and
\IEEEauthorblockN{Luca Benini}
\IEEEauthorblockA{\textit{Integrated Systems Laboratory, IIS} \\
ETH Zürich, Switzerland \\
lbenini@iis.ee.ethz.ch}
}

\maketitle

\begin{abstract}
Heterogeneous multi-core architectures combine a few ``host'' cores, optimized for single-thread performance, with many small energy-efficient ``accelerator'' cores for data-parallel processing, on a single chip. Offloading a computation to the many-core acceleration fabric introduces a communication and synchronization cost which reduces the speedup attainable on the accelerator, particularly for small and fine-grained parallel tasks.
We demonstrate that by co-designing the hardware and offload routines, we can increase the speedup of an offloaded DAXPY kernel by as much as 47.9\%. Furthermore, we show that it is possible to accurately model the runtime of an offloaded application, accounting for the offload overheads, with as low as 1\% MAPE error, enabling optimal offload decisions under offload execution time constraints.
\end{abstract}


\newacronym[plural=WANs, firstplural={Wide Area Networks (WANs)}]{wan}{WAN}{Wide Area Network}
\newacronym[plural=WSNs, firstplural={Wireless Sensor Networks (WSNs)}]{wsn}{WSN}{Wireless Sensor Network}
\newacronym{simd}{SIMD}{Single Instruction Multiple Data}
\newacronym{os}{OS}{Operating System}
\newacronym{ble}{BLE}{Bluetooth Low-Energy}
\newacronym{wifi}{Wi-FI}{Wireless Fidelity}
\newacronym[plural=DVS, firstplural={Dynamic Vision Sensors (DVS)}]{dvs}{DVS}{Dynamic Vision Sensor}
\newacronym{ptz}{PTZ}{Pan-Tilt Unit}
\newacronym[plural=FLLs,firstplural=Frequency Locked Loops (FLLs)]{fll}{FLL}{Frequency Locked Loop}
\newacronym{dram}{DRAM}{Dynamic Random Access Memory}
\newacronym{fpu}{FPU}{Floating Point Unit}
\newacronym{dma}{DMA}{Direct Memory Access}
\newacronym[plural=LUTs, firstplural={Lookup Tables (LUTs)}]{lut}{LUT}{Lookup Table}
\newacronym[plural=FPGAs, firstplural={Field Programmable Gate Arrays (FPGAs)}]{fpga}{FPGA}{Field Programmable Gate Array}
\newacronym{dsp}{DSP}{Digital Signal Processing}
\newacronym{mcu}{MCU}{Microcontroller Unit}
\newacronym{spi}{SPI}{Serial Peripheral Interface}
\newacronym{cpi}{CPI}{Camera Parallel Interface}
\newacronym{fifo}{FIFO}{First-In First-Out Queue}
\newacronym{uart}{UART}{Universal Asynchronous Receiver-Transmitter}
\newacronym{raw}{RAW}{Read-After-Write}
\newacronym[plural=ISAs, firstplural={Instruction Set Architectures (ISAs)}]{isa}{ISA}{Instruction Set Architecture}
\newacronym{xbar}{XBAR}{crossbar}
\newacronym[firstplural=Scratch-Pad Memories (SPMs)]{spm}{SPM}{Scratch-Pad Memory}
\newacronym{ppa}{PPA}{Power Performance Area}
\newacronym{ste}{STE}{Straight-Through-Estimator}
\newacronym[plural=PTUs, firstplural={Pan-Tilt Units}]{ptu}{PTU}{Pan-Tilt Unit}
\newacronym{mdf}{MDF}{Medium-density fibreboard}
\newacronym{cvat}{CVAT}{Computer Vision Annotation Tool}
\newacronym{coco}{COCO}{Common Objects in Context}
\newacronym{soa}{SoA}{State of the Art}
\newacronym{sf}{SF}{Sensor Fusion}
\newacronym{dl}{DL}{Deep Learning}
\newacronym{bn}{BN}{Batch Normalization}
\newacronym{FGSM}{FBK}{Fast Gradient Sign Method}
\newacronym{lr}{LR}{Learning Rate}
\newacronym{sgd}{SGD}{Stochastic Gradient Descent}
\newacronym{gd}{GD}{Gradient Descent}
\newacronym{sta}{STA}{Static Timing Analysis}
\newacronym[plural=GPIOs, firstplural={General Purpose Inupt Outputs (GPIOs)}]{gpio}{GPIO}{General Purpose Input Output}
\newacronym[plural=LDOs, firstplural={Low Dropout Regulators (LDOs)}]{ldo}{LDO}{Low Dropout Regulator}
\newacronym{inq}{INQ}{Incremental Network Quantization}
\newacronym{CV}{CV}{Computer Vision}
\newacronym{EoT}{EoT}{Expectation over Transformation}
\newacronym{RPN}{RPN}{Region Proposal Network}
\newacronym{TV}{TV}{Total Variation}
\newacronym{NPS}{NPS}{Non-Printability Score}
\newacronym{STN}{STN}{Spatial Transformer Network}
\newacronym{MTCNN}{MTCNN}{Multi-Task Convolutional Neural Network}
\newacronym{YOLO}{YOLO}{You Only Look Once}
\newacronym{SSD}{SSD}{Single Shot Detector}
\newacronym{SOTA}{SOTA}{State of the Art}
\newacronym{NMS}{NMS}{Non-Maximum Suppression}
\newacronym{ic}{IC}{Integrated Circuit}
\newacronym{rf}{RF}{Radio Frequency}
\newacronym{tcxo}{TCXO}{Temperature Controlled Crystal Oscillator}
\newacronym{jtag}{JTAG}{Joint Test Action Group industry standard}
\newacronym{swd}{SWD}{Serial Wire Debug}
\newacronym{sdio}{SDIO}{Serial Data Input Output}
\newacronym[plural=PCBs, firstplural={Printed Circuit Boards (PCB)}]{pcb}{PCB}{Printed Circuit Board}
\newacronym[plural=ASICs, firstplural={Application Specific Integrated Circuits}]{asic}{ASIC}{Application Specific Integrated Circuit}
\newacronym[plural=BNNs, firstplural={Binary Neural Networks (BNNs)}]{bnn}{BNN}{Binary Neural Network}
\newacronym[plural=NNs, firstplural={Neural Networks}]{nn}{NN}{Neural Network (NNs)}
\newacronym[plural=SCMs, firstplural={Standard Cell Memories (SCMs)}]{scm}{SCM}{Standard Cell Memory}
\newacronym{ann}{ANN}{Artificial Neural Networks}
\newacronym{ml}{ML}{Machine Learning}
\newacronym{ai}{AI}{Artificial Intelligence}
\newacronym{iot}{IoT}{Internet of Things}
\newacronym{fft}{FFT}{Fast Fourier Transform}
\newacronym[plural=OCUs, firstplural={Output Channel Compute Units (OCUs)}]{ocu}{OCU}{Output Channel Compute Unit}
\newacronym{alu}{ALU}{Arithmetic Logic Unit}
\newacronym{mac}{MAC}{Multiply-Accumulate}
\newacronym[firstplural={systems-on-chip (SoCs)}]{soc}{SoC}{system-on-chip}
\newacronym[firstplural={multi-processor systems-on-chip (MPSoCs)}]{mpsoc}{MPSoC}{multi-processor system-on-chip}
\newacronym{PGD}{PGD}{Projected Gradient Descend}
\newacronym{CW}{CW}{Carlini-Wagner}
\newacronym{OD}{OD}{Object Detection}
\newacronym{rrf}{RRF}{RADAR Repetition Frequency}
\newacronym{nlp}{NLP}{Natural Language Processing}
\newacronym{qam}{QAM}{Quadrature Amplitude Modulation}
\newacronym{rri}{RRI}{RADAR Repetition Interval}
\newacronym{radar}{RADAR}{Radio Detection and Ranging}
\newacronym{loocv}{LOOCV}{Leave-one-out cross validation}
\newacronym{bsp}{BSP}{Board Support Package}
\newacronym{ttn}{TTN}{The Things Network}
\newacronym{wip}{WIP}{Work in Progress}
\newacronym{json}{JSON}{JavaScript Object Notation}
\newacronym{qat}{QAT}{Quantization-Aware Training}
\newacronym{cls}{CLS}{Classification Error}
\newacronym{loc}{LOC}{Localization Error}
\newacronym{bkgd}{BKGD}{Background Error}
\newacronym{roc}{ROC}{Receiver Operating Characteristic}
\newacronym{frr}{FRR}{False Rejection Rate}
\newacronym{eer}{EER}{Equal Error Rate}
\newacronym{snr}{SNR}{Signal-to-Noise Ratio}
\newacronym{flop}{FLOP}{Floating-Point Operation}
\newacronym{fp}{FP}{Floating-Point}
\newacronym{fps}{FPS}{Frames Per Second}
\newacronym{gsc}{GSC}{Google Speech Commands}
\newacronym{mswc}{MSWC}{Multilingual Spoken Words Corpus}
\newacronym{demand}{DEMAND}{Diverse Environments Multichannel Acoustic Noise Database}
\newacronym[plural=SNNs, firstplural={Spiking Neural Networks (SNNs)}]{snn}{SNN}{Spiking Neural Network}
\newacronym[plural=DNNs, firstplural={Deep Neural Networks (DNNs)}]{dnn}{DNN}{Deep Neural Network}
\newacronym[plural=TCNs,firstplural=Temporal Convolutional Networks]{tcn}{TCN}{Temporal Convolutional Network}
\newacronym[plural=CNNs,firstplural=Convolutional Neural Networks (CNNs)]{cnn}{CNN}{Convolutional Neural Network}
\newacronym[plural=TNNs,firstplural=Ternarized Neural Networks]{tnn}{TNN}{Ternarized Neural Network}
\newacronym{ds-cnn}{DS-CNN}{Depthwise Separable Convolutional Neural Network}
\newacronym{rnn}{RNN}{Recurrent Neural Network}
\newacronym{gcn}{GCN}{Graph Convolutional Network}
\newacronym{mhsa}{MHSA}{Multi-Head Self Attention}
\newacronym{crnn}{CRNN}{Convolutional Recurrent Neural Network}
\newacronym{clca}{CLCA}{Convolutional Linear Cross-Attention}
\newacronym{bf}{BF}{Beamforming}
\newacronym{anc}{ANC}{Active Noise Cancellation}
\newacronym{agc}{AGC}{Automatic Gain Control}
\newacronym{se}{SE}{Speech Enhancement}
\newacronym{mct}{MCT}{Multi-Condition Training}
\newacronym{mcta}{MCTA}{Multi-Condition Training \& Adaptation}
\newacronym{pcen}{PCEN}{Per-Channel Energy Normalization}
\newacronym{mfcc}{MFCC}{Mel-Frequency Cepstral Coefficient}
\newacronym{asr}{ASR}{Automated Speech Recognition}
\newacronym{kws}{KWS}{Keyword Spotting}
\newacronym{odl}{ODL}{On-Device Learning}

\section{Introduction}

In the post Dennard's scaling era, heterogeneous computing offers a solution to improve system performance, by integrating different compute units tailored to a diverse range of applications onto a single system. This work targets a class of heterogeneous computing systems referred to as heterogeneous (or asymmetric) \glspl{mpsoc}, which have met increasing interest in recent years \cite{ignjatovic2022, ditzel2022, manticore}. 
These systems couple one or more high-performance ``host'' cores, guaranteeing high single-thread performance, with a multi-cluster fabric of energy-efficient ``accelerator'' cores, enabling the execution of intensive data-parallel computations at high energy-efficiency and performance points.

In this context, we refer to the process of handing over parts of the computation to the accelerator as \textit{job offloading}.
It is the programmer's responsibility to define the workload partition between the host and the accelerator, i.e. to define jobs to offload to the accelerator, and making a correct offload decision is non-intuitive \cite{che2009}. This decision is not only about determining \textit{if} a portion of the workload can benefit or not from offloading, but also about the specifics on \textit{how} to offload the workload, e.g. how many cores to employ for a job, which may have a significant impact on performance \cite{araujo2023}.

To complicate the matter, offloading introduces several overheads, in the form of synchronization and communication between host and accelerator, which add up to the runtime and energy consumption of the job execution on the
accelerator. Therefore, to enable effective fine-grained heterogeneous execution and exploit the full potential of heterogeneous architectures, these overheads must be reduced to a minimum.

Several works looked into the problem of quantifying offload overheads, but do not propose a generalized model, fail to assess the accuracy of the proposed models \cite{pei2016}, or both \cite{pei2016-2, che2009, kannan2015}.
All of these works study discrete CPU-GPU heterogeneous architectures, where CPU and GPU reside on separate chips interconnected by a PCIe bus. Due to the proprietary closed-source nature of these architectures, precise evaluation of the offload overheads' magnitude is not possible.
Finally, to the best of our knowledge, no prior work looked into reducing the overheads associated to offloading in heterogeneous \glspl{mpsoc}.

With this work, we show that by co-designing the hardware and offload routines it is possible to 1) decrease the offload overheads and improve offloaded application performance and 2) develop an accurate runtime model which can be used to formulate the offload decision as an optimization problem. 

\section{Implementation}

We develop this study on the Manticore \gls{mpsoc} \cite{manticore}, as it combines the benefits of recent architectural trends with a fully open-source hardware design, allowing for a complete understanding of the offload overhead cycles.
Manticore is a heterogeneous \gls{mpsoc} targeting data-parallel floating-point workloads.
It features a CVA6 host core \cite{zaruba2019} coupled to a symmetric multi-processor accelerator comprising a configurable number of energy-efficient cores, up to 288 in our experiments. The accelerator cores are organized in clusters of 9 cores each.

We extend Manticore's interconnect and CVA6's load-store unit and memory subsystem to support multicasting data from CVA6 to the individual clusters. Design details and implementation results are omitted as in this brief we focus on assessing and modeling the perfomance benefits of multicast communication on offload performance.

In addition, we designed and integrated a dedicated unit to handle the accelerator to host synchronization at the end of an offload. This unit implements a centralized credit counter. Upon an offload, CVA6 sets the number of accelerator clusters selected for offload as a threshold for the counter. When a cluster is done with the job, it atomically increments the counter by writing to a register which triggers the increment as a side effect. As soon as the counter reaches the threshold value set by CVA6, it automatically fires an interrupt notifying CVA6 of job completion.

\section{Results}
\label{sec:results}

\begin{figure}[t]
    \vspace{-2.7ex}
    \centering
    \subfloat{
        \includegraphics[width=0.23\textwidth]{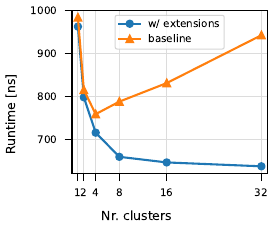}
    }
    \subfloat{
        \includegraphics[width=0.23\textwidth]{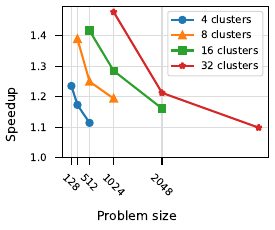}
    }
    \caption{Runtime of a 1024-dimension DAXPY job for various numbers of clusters employed (left); speedup of the DAXPY job with our extensions over the baseline implementation for various problem sizes and numbers of clusters (right).}
    \label{fig:comparison}
    \vspace{-1em}
\end{figure}

All experiments are conducted through cycle-accurate RTL simulations of the Manticore \gls{mpsoc} using QuestaSim 2022.3. The testbench drives all clocks at a frequency of \SI{1}{\giga\hertz}, thus all runtimes reported in nanoseconds are in 1:1 correspondence with the runtime in CPU cycles.

To quantify the impact of the offload overheads, and evaluate the improvement from our extensions, we measure the runtime of an offloaded 1024-dimension DAXPY kernel with and without our extensions. \cref{fig:comparison} (left) presents the results of this comparison. The runtime in the baseline implementation presents a global minimum, owing to the fact that the offload overheads actually increase with the number of clusters. In fact, the job handler and arguments have to be dispatched to every cluster and, as this can only be done sequentially, the overhead depends linearly on the number of clusters. In the baseline design, when the number of clusters in the accelerator grows above four, the offload overhead starts to dominate (as it grows linearly, while the amount of work per cluster decreases).

With the multicast extension, dispatching occurs in parallel, so the overhead is instead constant. Thus, we can leverage additional clusters up to 32 while still decreasing execution time. The improvement of the proposed multicast solution reaches more than 300 cycles difference in the 32-clusters configuration. Offloading to more clusters would lead to negligible further improvements because of Amdahl's law. 

In the previous experiment, we used a fixed problem size to highlight the dependency with the number of clusters, serving to show that \textit{optimizing the offload overheads is most critical for large many-core accelerators}, as the workload distributed to each cluster gets smaller. \cref{fig:comparison} (right) adds the dependency on the problem size to the picture. It displays the speedup with our extensions over the baseline for different problem sizes and numbers of clusters. The speedup is always greater than one, although, for a fixed number of clusters employed, it decreases with the problem size. Indeed, as the problem size increases, also the job execution time increases, while the time to communicate the job information stays constant. It follows that the improvement on the latter has a smaller impact on the runtime (and therefore the speedup) of the overall offload. These results confirm that \textit{optimizing the overheads is most significant for fine-grained parallel tasks}.

By inspecting the hardware and the compiled application, we can quantitatively model the overall runtime of an offloaded DAXPY kernel of dimension $N$ onto $M$ clusters:
\begin{equation}
    \label{eq:runtime}
    \mathit{\hat{t}_{offl}}(M, N) = 367 + \frac{N}{4} + \frac{2.6*N}{M*8}
\end{equation}
According to this model, the offloaded DAXPY kernel responds to Amdahl's law. The speedup attained by scaling the computation to multiple accelerator clusters is limited by the serial fraction made up in part by the offload overheads.

We validate the accuracy of the model on multiple problem sizes and number of clusters. For each problem size ($N \in \{256, 512, 768, 1024\}$) we calculate the mean absolute percentage error (MAPE) over all tested offload configurations ($M \in \{1, 2, 4, 8, 16, 32\}$):
\begin{equation}
    \mathit{MAPE}(N) = \frac{100}{M} * \sum_{M} \frac{\lvert \mathit{t_{offl}}(M, N) - \mathit{\hat{t}_{offl}}(M, N) \rvert }{\mathit{t_{offl}}(M, N)}
\end{equation}
The error is consistently lower than 1\%, proving that the model can accurately estimate offloaded application runtime.

Such a model can be used to solve the offload decision problem for $M$. For instance, under the maximum runtime constraint $\mathit{t_{offl}}(M) \leq \mathit{t_{max}}$, we can easily invert \cref{eq:runtime} to derive the minimum number of clusters that would be needed for the offload to satisfy the constraints:
\begin{equation}
    M_{min} = \left \lceil{\frac{2.6*N}{8*(t_{max} - 367 - \frac{N}{4})}}\right \rceil
\end{equation}

\section{Conclusion}
\label{sec:conclusion}

We showed how, by enabling multicasting in the host-to-accelerator clusters interconnect, we can improve the speedup of an offloaded application by as much as 47.9\%, as measured on a low vector dimension (1024) DAXPY kernel.
Furthermore, we demonstrated the feasibility of developing an accurate runtime model accounting for the offload overheads, and showed how such a model can be used to derive optimal offloading parameters under an offload execution time constraint.

\bibliographystyle{IEEEtranS}
\bibliography{refs}

\end{document}